# Ultralow-power all-optical switching via a chiral Mach-Zehnder interferometer

Yaping Ruan[1,4], Haodong Wu[1,4], Shijun Ge[1], Lei Tang[1], Zhixiang Li[1], Han Zhang[1], Fei Xu[1], Wei Hu[1], Min Xiao[1,2], Yanqing Lu[1*] and Keyu Xia[1,3*]

[1]National Laboratory of Solid State Microstructures, Key Laboratory of Intelligent Optical Sensing and Manipulation, College of Engineering and Applied Sciences, and Collaborative Innovation Center of Advanced Microstructures, Nanjing University, Nanjing 210023, China

[2]Department of Physics, University of Arkansas, Fayetteville, Arkansas 72701, USA

[3]Jiangsu Key Laboratory of Artificial Functional Materials, Nanjing University, Nanjing 210023, China

[4]These authors contributed equally: Yaping Ruan, Haodong Wu

*Correspondence and requests for materials should be addressed to Y.L. (email: yqlu@nju.edu.cn) or K.X. (email: keyu.xia@nju.edu.cn)

**All-optical switching increasingly plays an important role in optical information processing. However, simultaneous achievement of ultralow power consumption, broad bandwidth and high extinction ratio remains challenging. We experimentally demonstrate an ultralow-power all-optical switching by exploiting chiral interaction between light and optically active material in a Mach-Zehnder interferometer (MZI). We achieve switching extinction ratio of 20.0(3.8) and 14.7(2.8) dB with power cost of 66.1(0.7) and 1.3(0.1) fJ/bit, respectively. The bandwidth of our all-optical switching is about 4.2 GHz. Our theoretical analysis shows that the switching bandwidth can, in principle, exceed 110 GHz. Moreover, the switching has the potential to be operated at few-photon level. Our all-optical switching exploits a chiral MZI made of linear optical components. It excludes the requisite of high-quality optical cavity or large optical nonlinearity, thus greatly simplifying realization.**

**Our scheme paves the way towards ultralow-power and ultrafast all-optical information processing.**

Optical switching is a key component in many optical technologies such as optical communication networks. It is also the building block in realizing optical logic device, optical computation and the emerging field of photon-based artificial intelligent. In many cases, the performance of a whole system is limited by optical switching.

Realizing fast and ultralow-power optical switching with a usable extinction ratio is highly desired but extremely challenging. Optical switching has been demonstrated by using thermal-optical effect[1,2], liquid crystal optical material[3,4], magneto-optical effect[5,6], acousto-optical effect[7,8], and electro-optical effect[9–13]. Among these kinds, electro-optical switching with a Mach-Zehnder interferometer (MZI) is widely used in optical communication[11–13]. However, the power cost, which is usually defined as the control light power for switching the signal or the consumption of energy per control light pulse, in such electro-optical switching is typically high because it requires optical-electrical/electrical-optical conversion. In stark contrast, all-optical switching exhausts much lower energy and thus attracts intensive studies[14–19]. For a practical application, ultralow power consumption, broad bandwidth and high extinction ratio are the most important performance of an optical switching. In contrast to other mechanisms, all-optical switching has the potential to meet all these three requisites at the same time, but is yet to demonstrate this capability.

In recent years, high-speed, all-optical switching operating at low power level have been realized in various platforms[20–49]. One of the most popular methods involves modifying the refractive index contrast of a Kerr nonlinear medium with a strong pump laser beam[20]. However, it is challenging to achieve a nonlinearity-based all-optical switching with ultralow power cost and high extinction ratio, because optical Kerr nonlinearity in most materials is typically too weak. This weak nonlinearity constrains the practical availability of all-optical switching enormously. Two schemes have been proposed to solve this problem. One scheme is based on quantum-interference, such as electromagnetically induced transparency (EIT) in which linear susceptibility vanishes and the nonlinear interaction strength can be greatly enhanced by many orders of magnitude[21–38]. However, the quantum-interference-enhanced large nonlinear optical coefficient also results in narrow switching bandwidth and subsequently slows down switching speed[39]. The other scheme uses high-quality microcavity to enhance the nonlinear photon-photon interaction[40–45]. But a high-quality cavity is difficult to fabricate and it also dramatically reduces the switching bandwidth. Besides, there are other schemes to realize all-optical switching, such as optical bistability[46], electromagnetically induced absorption[47], three-wave mixing[48] and coulomb blockade[49].

Chiral light-matter interaction has been used to realize striking optical technologies such as optical nonreciprocity[50-62] and chiral quantum information processing[63-65]. Here, we experimentally demonstrate an ultralow-power all-optical switching by exploiting chiral interaction between light and an optical material with optical activity in a MZI. We achieve all-optical switching with an extinction ratio up to 20 dB by using a femtojoule-level weak control light. Theoretically, the bandwidth of the optical switching can exceed one hundred of gigahertz, allowing ultrafast operation with about 10 ps switching speed. Our work opens a door to conduct ultrafast and ultralow-power optical information processing.

## Results

**Schematic of concept.** Our key idea of ultralow-power all-optical switching using a chiral MZI is schematically illustrated in Fig. 1a. Our all-optical switching crucially relies on the chiral interaction between light and a quartz crystal with optical activity in a MZI. The material is “chiral” in the sense of its different response to left and right circular polarized (LCP and RCP) light. The material with optical activity has different refractive indices for LCP and RCP light, causing different phase shifts to these two orthogonal polarized light beams. This phase shift difference is also proportional to the material length. Thus, similar to a Faraday rotator, by choosing a proper length, the chiral material can convert a Horizontally-polarized (H-polarized) light to Vertically-polarized (V-polarized) without using an external magnetic field.

By using a V-polarized control light, we can adjust the polarization of light beam propagating in the chiral material to realize an all-optical switching. The mechanism is following. A beam splitter is used to mix the H-polarized signal and the V-polarized control beams. The mixed light beams are then sent to two arms of the chiral MZI. In our schematic, the polarization of the light beam introduced to the MZI can be adjusted by the control light. In the presence of the V-polarized control light, namely for the logic "off" state of the control, the light beams outcome from the first beam splitter (BS1) are RCP. This RCP beam in the upper arm of the MZI transmits through the chiral material and then is subject to a phase shift but remains its polarization. We choose the relative phase mod($\Delta\varphi, 2\pi$)=0 between two arms of the MZI in an ideal case for this RCP beam. In this case, the upper-port from the MZI is "dark" due to the destructive interference on the second beam splitter (BS2), corresponding to the signal "off" state. In the absence of the control light, i.e. the logic “on” state of the control, the light after BS1 only includes the H-polarized component from the signal. After passing through the chiral material, this H-polarized field in the upper arm becomes V-polarized, avoiding interference with the H-polarized field in the lower arm on BS2. As a result, the outcome intensity from the upper port is high, indicating an “on”-state signal output. Clearly, the signal light can be switched “on” or “off” by the control light without the need of power-consuming modulation of the MZI or material as usual. On the other hand, we only use a cavity-free optical system and linear chiral material. This design allows us to switch the signal at an ultrafast speed.

**Experimental setup.** Figure 1b shows the experimental setup realizing the aforementioned schematic idea. Our experiment is based on a chiral MZI in which a material with optical activity is embedded in one arm. Our chiral MZI is essentially distinct in underlying physics from a conventional MZI. With a proper length of the chiral material, H-polarized and RCP light beams entering the MZI with fixed-length arms will have very different outcomes from the output port detected by an APD (Model APD440A, Thorlabs). Below we will explain in detailed how we experimentally realize all-optical switching by using this chiral MZI. A laser beam from a tunable external cavity diode laser (Model DLC pro, Toptica Company) is split into two beams. This laser has a very narrow linewidth less than 100 kHz. The upper beam acts as the signal light. The lower beam is chopped into pulses. It is then divided into two parts with equal power by a 50:50 BS: one plays as the role of the control light, while the other works as a reference light to monitor the laser power. The power of the reference light is monitored by the APD on the left. By using a pair of QWP and HWP, the signal and control lights are prepared in H- and V-polarized, respectively. The relative phase of the signal and control beams is fixed to $\pi/2$ using an electric controlled liquid crystal retarder (LCR). The signal and control beams are then merged into one beam by a polarization beam splitter (PBS). In our arrangement, this combined beam is RCP or H-polarized when the control light is present or absent, corresponding to “off” or “on” logic state, respectively. Note that the control and signal light powers are equal in the presence of the control beam. The combined light beam is then coupled into a single-mode polarization-maintaining fiber and is reshaped to a transversal Gaussian mode. After mode reshaping, it enters the chiral MZI. The chiral material is inserted in the upper arm of the MZI. When the control field is in the logic “off” state, i.e. in the presence of the control light, the beam entering the MZI is RCP. The chiral material only causes a phase shift to light in the upper arm but keeps its polarization unchanged. We adjust the relative phase $\Delta\varphi$ between the two arms of MZI to be as small as possible by a variable delay stage. In the ideal case of $\Delta\varphi = 0$, light beams in two arms destructively interfere on the second BS2 and an “off” state output of signal with vanishing intensity is obtained. To switch on the signal, we set the control field in logic “on” state. In this case, the light beam entering the MZI is H-polarized. The upper beam is then converted to V-polarized by the chiral material. Thus, the upper and lower beams transmit through BS2 independently without interference.

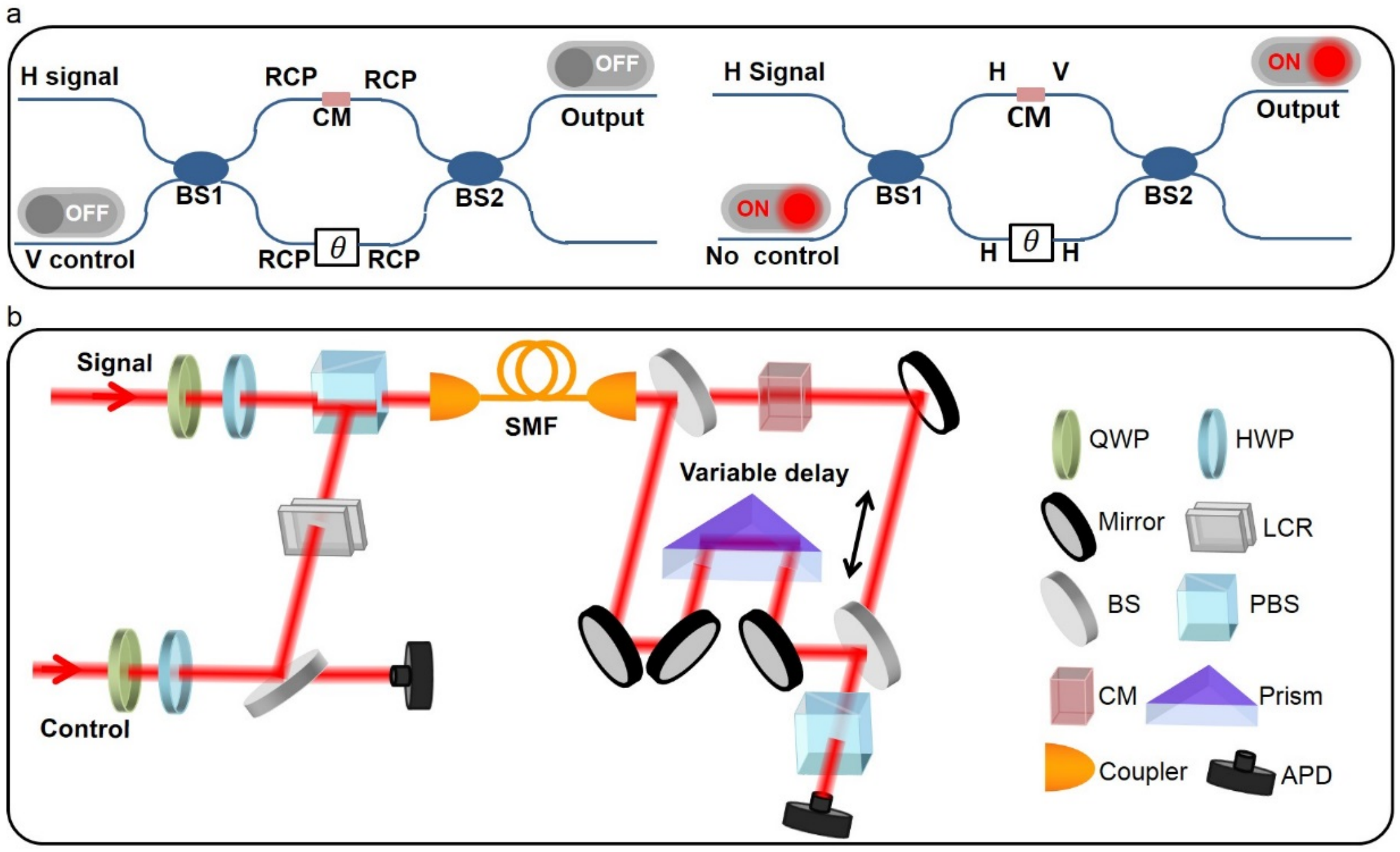

**Fig. 1 Schematic and experimental set-up of all-optical switching. a** Schematic of all-optical switching using a chiral Mach-Zehnder interference (MZI). The signal light is horizontally (H)-polarized and the control light is vertically (V)-polarized. The chiral material with optical activity causes a phase to a circularly-polarized light beam but converts a H-polarized field to H-polarized. The control light can switch "on" and "off" the signal light. The effective phase shifter $\theta$ is used to compensate the unwanted phase difference in the two MZI arms. **b** Experimental implementation of all-optical switching depicted in **a**. It includes preparation of signal and control light, MZI and chiral material. The phase difference between two arms of interferometer is adjusted by a variable delay stage. A chiral quartz crystal is placed in one arm of the MZI to transform the polarization and phase of the beam. The output signal is detected by an avalanche photodiode (APD). Details of all optical devices: QWP, quarter wave-plate; HWP, half wave-plate; Mirror, reflecting mirror; LCR, Liquid crystal retarder; BS, beam splitter; PBS, polarizing beam splitter; CM, chiral material; Coupler, fiber coupler; APD, avalanche photodiode.

The outcoming signal is then in "on" state with a high-level output intensity. In practical case, the phase difference $\Delta\varphi$ can be non-zero but small. As a result, the "off"-state output signal is non-vanishing but at a low level. At the same time, the detected output power of an "on"-state signal is lower than the value of a perfect arrangement. The measured extinction ratio will therefore reduce due to this experimental imperfection. On the other hand, noise from electronic circuits and background will also cause the decrease of extinction ratio and the signal-to-noise ratio (SNR).

**Theoretical performance of all-optical switching.** Now we present a theoretical model to analysis the extinction ratio and bandwidth of our all-optical switching. In the presence of the control light, the output signal intensity with a harmonic frequency $\nu$ is given by

$$I_{\mathrm{off}}(\nu) = \frac{I_0}{2}\left[\eta_1 + \eta_2 - 2\sqrt{\eta_1\eta_2}\cos(\frac{2\pi\Delta L}{c}\nu)\right], \tag{1}$$

where $I_0$ is the total intensity of the input signal light. Note that the intensity of control light is equal to that of the input signal light. $\eta_1$ and $\eta_2$ are the total effective transmittances of light beams in the lower and upper paths, respectively. Here, we already consider the loss and absorption during light propagation and the imperfection of the BS. $\Delta L$ is the effective optical path difference (OPD) of two MZI arms, causing a relative phase $\Delta\varphi(\nu) = 2\pi\Delta L\nu/c$. To avoid the nonzero "off"-state output in the ideal case, we need mod[$\Delta\varphi(\nu), 2\pi$]=0 that $\Delta\varphi(\nu)$ is the integer times of $2\pi$. However, the OPD $\Delta L$ is nonzero due to experimental imperfection. Thus, we adjust the length of two MZI arms to guarantee $\Delta\varphi(\nu)$ as small as possible for the central frequency of the light. Without applying the control light, the laser beams in two arms are orthogonal in polarization. Therefore, the destructive interference on the second BS disappears. As a result, the output signal field is switched to the "on" state. The output signal intensity is given by

$$I_{\mathrm{on}} = \frac{I_0}{4}\eta_1. \tag{2}$$

The "on" state output is independent of light wavelength because the absence of destructive interference. The extinction ratio is evaluated as

$$R(\nu) = -10\log_{10}\frac{I_{\mathrm{off}}(\nu)}{I_{\mathrm{on}}} \ . \tag{3}$$

We set $2\pi\Delta L\nu_0/c = 2n\pi$ ($n$ is an integer) such that mod[$\Delta\varphi(\nu_0)$,$2\pi$] $=0$ for the central frequency $\nu_0$. In this case, the output intensity of the "off"-state signal light is low and given by

$$I_{\mathrm{off}}(\nu_0) = \frac{I_0}{2}\left(\sqrt{\eta_1} - \sqrt{\eta_2}\right)^2. \tag{4}$$

Obviously, in an ideal arrangement of $\eta_1 = \eta_2$, the output signal light can be completely switched "off" to zero. When the light frequency is $\nu_0$, we obtain the maximal extinction ratio

$$R_{\max} = -10\log_{10}\left[2\left(1-\sqrt{\xi}\right)^2\right], \tag{5}$$

with $\xi = \eta_2/\eta_1$.

Below we analysis the switching bandwidth. When the light wavelength varies to $\nu \neq \nu_0$, the value mod$[\Delta\varphi(\nu), 2\pi]$ is nonzero but remains small. The output signal intensity then becomes

$$I_{\text{off}}(\nu) = \frac{I_0}{2}\left[\eta_1 + \eta_2 - 2\sqrt{\eta_1\eta_2}\cos\left(\frac{2\pi\Delta L}{c}\Delta\nu\right)\right], \quad (6)$$

with $\Delta\nu = \nu - \nu_0 \ll \nu_0$. Note that $2\pi\Delta L\nu_0/c = 2n\pi$, the extinction ratio is then expressed as

$$R = -10\log_{10}2\left[1 + \xi - 2\sqrt{\xi}\cos\left(\frac{2\pi\Delta L}{c}\Delta\nu\right)\right]. \quad (7)$$

The 3 dB bandwidth defined as the full width at half maximum (FWHM) is

$$BW = \frac{c}{\pi\Delta L}\arccos\left[\frac{1+\xi-\left|1-\sqrt{\xi}\right|/\sqrt{2}}{2\sqrt{\xi}}\right]. \quad (8)$$

According to Eq. 8, we can see that the bandwidth of our all-optical switching can be, in principle, very broad if $\Delta L \to 0$ because the material is assumed linear and the system excludes the use of an optical cavity. The chirality of material is also dependent on light wavelength. Therefore, the operating bandwidth relies on the dispersion of material. On the other hand, the bandwidth is crucially limited by the OPD $\Delta L$ in experiment.

**All-optical switching with a relative high control power.** We first demonstrate an all-optical switching with relative high control power. The control and signal light beams have the same frequency of 384.2793 THz. In the plane of measurement screen, we place two APDs shown in Fig. 1b to record the power of the control light and output signal light, respectively. The APDs are connected to an oscilloscope to convert the optical signals into electric signals. The control light is modulated at 1.8 kHz by a mechanical chopper wheel. The power of the control light is 238.3(2.5) pW, corresponding to a power cost of 66.2(0.7) fJ/bit. We measure the time-dependent power of control light (hollow red circle) and output signal light (hollow blue square) as shown in Fig. 2. The single-measurement extinction ratio of optical switching can reach 20.0(3.8) dB. Clearly, we achieve a sub-nanowalt optical switching with a high extinction ratio. Note that a lower power cost is available because our optical switching only uses linear material. Below, we demonstrate an ultralow-power optical switching by fixing the light modulation rate at 1.8 kHz but changing the control power.

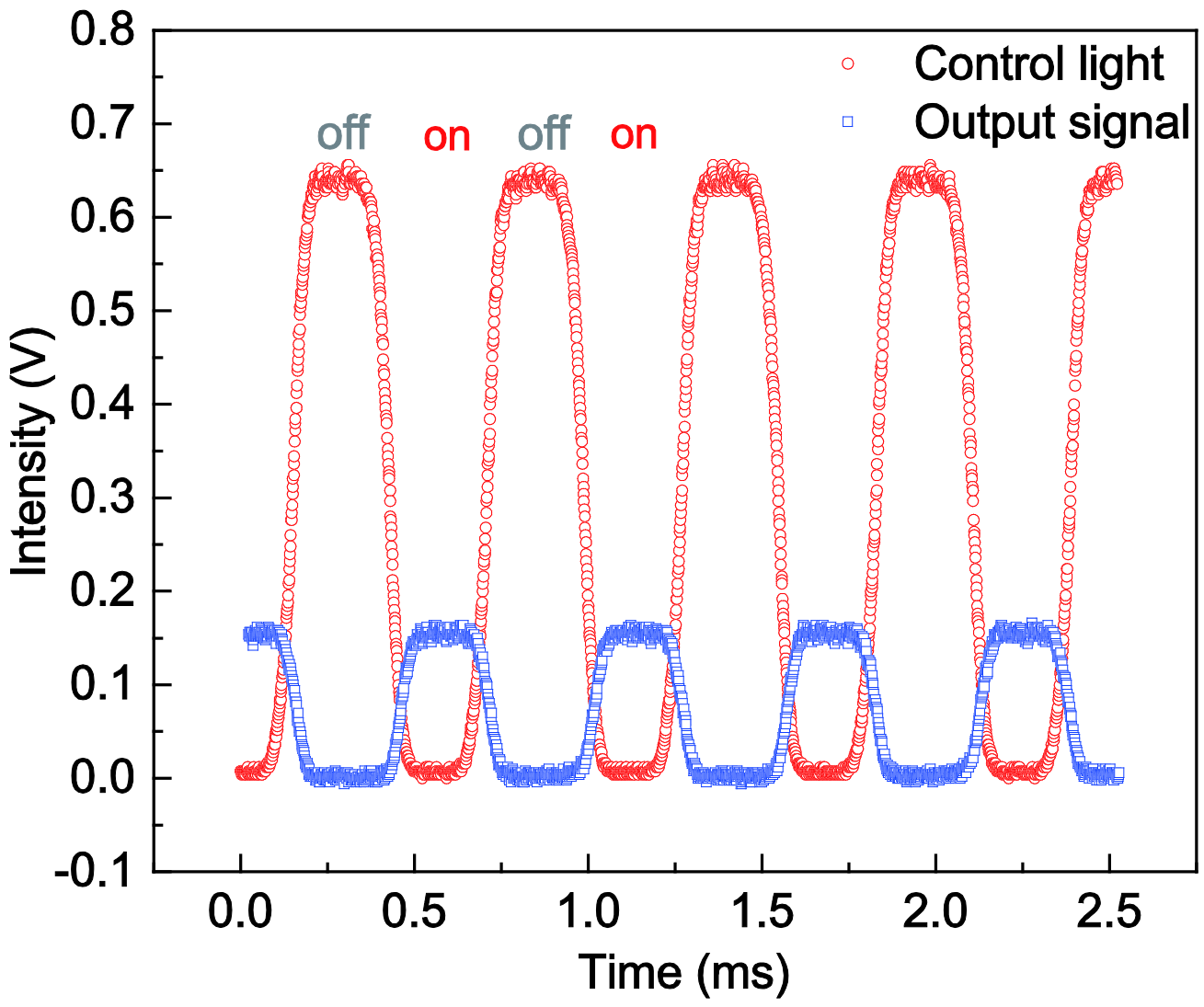

**Fig. 2 Experimental demonstration of all-optical switching with 238.3(2.5) pW control light pulses.** Time-dependent control light intensity is represented by hollow red circle and output signal light intensity is represent by hollow blue square. The frequency of the control light and signal light is 384.2793 THz. The "on" state control and output signal light powers are 238.3(2.5) pW and 57.8(1.5) pW, respectively.

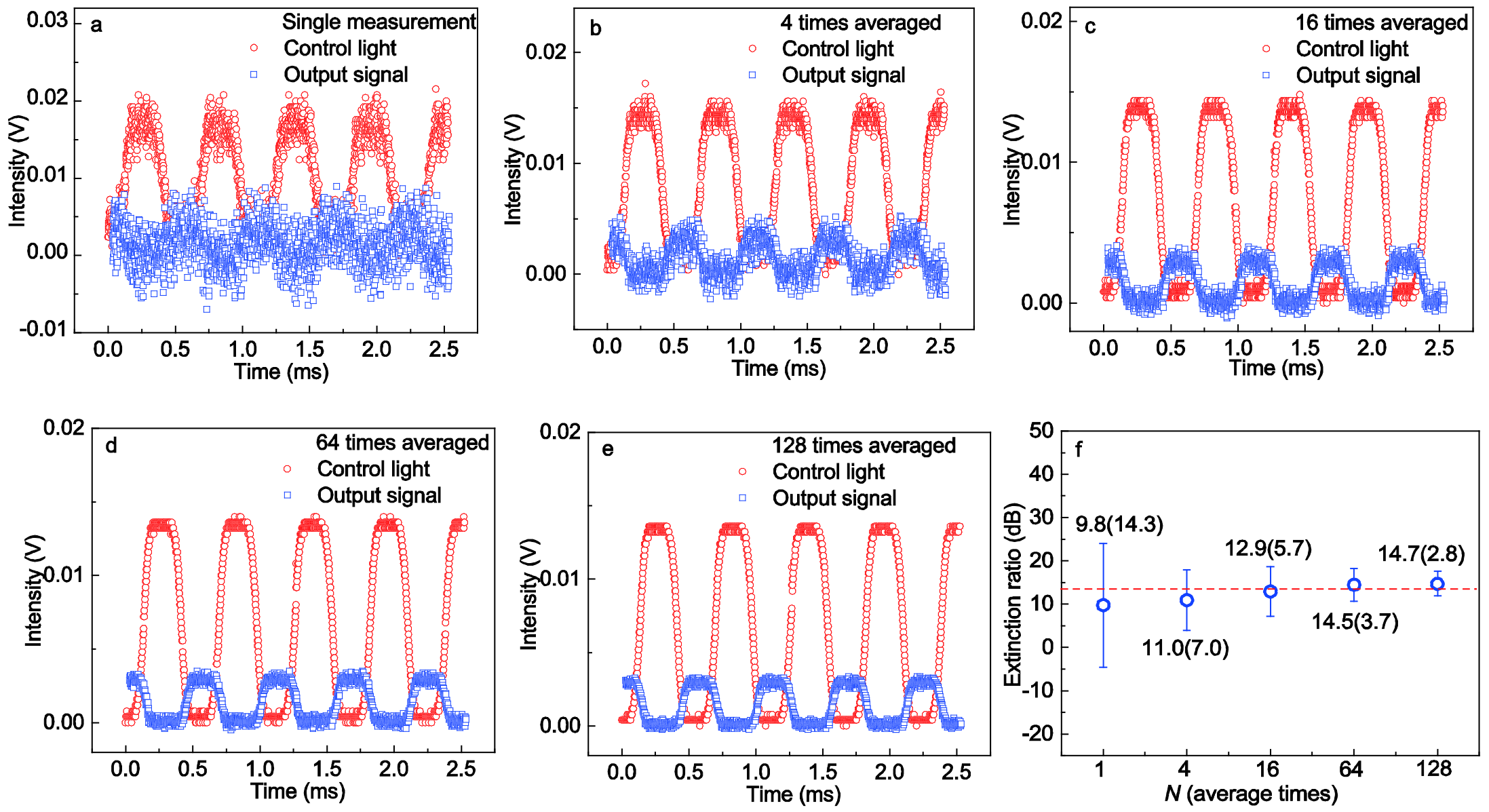

**Fig. 3 Ultralow-power all-optical switching with only 1.3(0.1) fJ/bit power cost. a-e,** results of single measurement, 4 times measurements averaged, 16 times measurements averaged, 64 times measurements averaged, and 128 times measurements averaged, respectively. **f,** Corresponding extinction ratio of all-optical switching in **a-e**.

**All-optical switching with a femtojoule-level power cost.** In Fig. 3, we demonstrate an ultralow-power optical switching with a control light power as low as system noise level. In experiment, we reduce the power of control light and signal light to 4.5(0.5) pW. The energy of each control laser pulse decreases down to 1.3(0.1) fJ accordingly. It can be clearly seen from Fig. 3a that the signal light can be effectively switched on and off. In this single-measurement case, the SNR is low because the energy of both the signal and control light pulses is just higher than system noise. The system noise mainly comes from electronic noise. If system noise can be reduced, we can considerably improve the SNR of the output signal. To demonstrate this possibility, we average the output signals many times in oscilloscope. Obviously, the influence of electronic noise is reduced and the SNR is considerably improved as averaging times increases, see figs. 3b-e. The switching of the output signal also becomes clearer. Because the effective noise reduces with the averaged times increasing, the extinction ratio is improved. As shown in Fig. 3f, the average extinction ratio is 9.8(14.3) dB, 11.0(7.0) dB, 12.9(5.7) dB, 14.5(3.7) dB and 14.7(2.8) dB for single measurement, 4 times averaged, 16 times averaged, 64 times averaged and 128 times averaged, respectively. The large error of extinction ratio in the cases of small average times results from large electronic noise. For *N*-time average, the error can be roughly suppressed by a factor of $\sqrt{N}$ according to measurement theory. Thus, the error range of extinction ratio continuously reduces to a small level and the extinction ratio is gradually improved to a saturated value 14.7(2.8) dB with the averaging time increasing. The extinction ratio of 14.7(2.8) dB approaches to the best performance of a noise-free system. Therefore, we can realize an all-optical switching even though the power cost is as low as 1.3(0.1) fJ/bit. If system noise can be reduced, an extinction ratio of about 14.7 is available. Such value of extinction ratio is already high for a femtojoule-level all-optical switching.

**System noise**. In practice, there is always some system noise, for example from electronic circuits, entering the detector. This noise reduces the extinction ratio and limits the minimal usable control light power switching the signal. We assume a system noise of power level $P_n$, yielding a noise intensity $I_n$ to the detector. When the outcoming signal is in the off state, the detected light intensity is $I_n + I_{off}(\nu)$. If the signal is switched on, then the light intensity of $I_n + I_{on}$ will be detected. We assume that the light intensity $I_0$ corresponds to a light power $P_c$. Then, the extinction ratio at frequency $\nu_0 = \nu$ when taking into account system noise is given by

$$R = -10\log_{10}\frac{2P_n+P_c\left(\sqrt{\eta_1}-\sqrt{\eta_2}\right)^2}{2P_n+P_c\eta_1/2} \quad . \tag{9}$$

To validate this power dependence of the extinction ratio and identify the system noise level, we have measured the extinction ratios in 8 different control power costs, as shown in Fig. 4. It can be seen that the extinction ratio first rapidly increases with the power cost of control light in noise-power level and then gradually becomes saturate when the power of control light grows to a high level. In addition, we have theoretically fitted experimental data according to Eq. 9. During the fitting, $\eta_1 = 0.98$, $\eta_2 = 0.82$. We can see from Fig. 4 that the theoretical fitting is in good agreement with experimental data. The fitting parameter $P_n$ is 0.20(0.09) pW. Therefore, the overall system noise in our system is about 0.20(0.09) pW, corresponding to 218 photons within a signal output pulse. According to our understanding, the system noise mainly comes from the APD.

Evaluated from the noise equivalent power (3.5 fW/$\sqrt{\mathrm{Hz}}$) of the APD and the measurement bandwidth (~4.3 kHz), the minimum detectable power is about 0.23 pW, corresponding to 0.06 fJ/bit in our experiment. Thus, the value of $P_n$ is in agreement with noise level of measurement devices provided by manufactories.

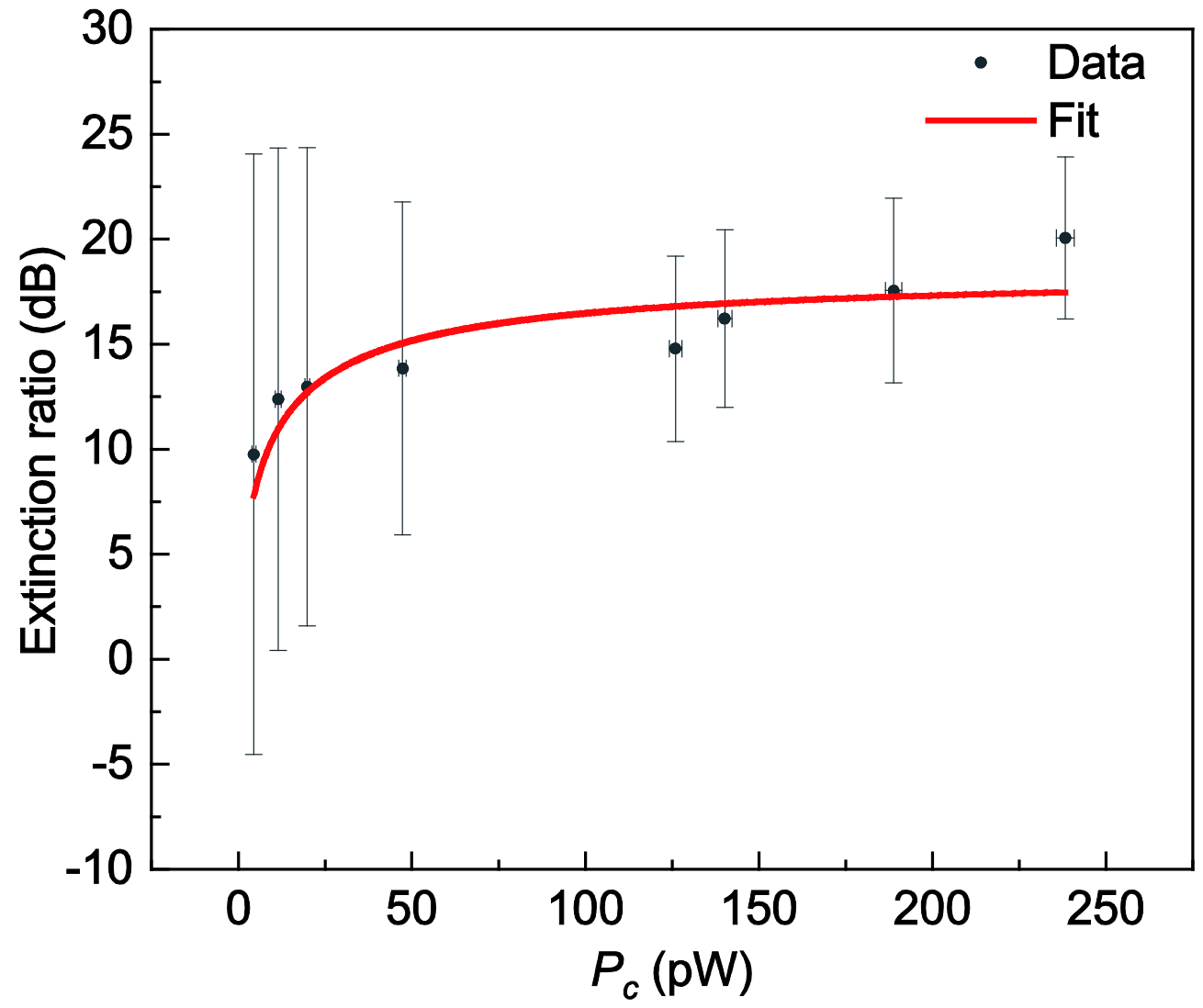

**Fig. 4 Extinction ratios for different control light powers.** Black circles represent the experimental results. The red curve shows the theoretical fitting.

According to the above results, to make the SNR exceed one ($I_{on} > I_n$) for practical use, the power cost of control light needs to be greater than 0.23 fJ/bit according to Eq. 2. However, the minimum power cost of control light in our all-optical switching is 1.3(0.1) fJ/bit. The reason is as follows: in our experimental measurements, the minimum usable power cost of control light is also limited by the measuring precision and minimal trigger level of oscilloscope. If the power of control light was further reduced, the measurement signals will be difficult to trigger and the measuring error will be very large. If system noise can be reduced and measuring precision of oscilloscope can be improved, an all-optical switching with smaller power cost will be obtainable.

**Bandwidth of all-optical switching.** The switching bandwidth is another important feature clarifying the performance of optical switching. It determines how fast one can switch on and off the signal. We evaluate the bandwidth of our optical switching by experimentally measuring the extinction ratio for different light frequencies (see blue circles Fig. 5a). To obtain the bandwidth, we calculate the FWHM of extinction ratio according to the measurements of the on- and off-state output signal intensities at 25 different frequencies. The optimal frequency $v_0$ is 384.2782 THz corresponding to minimal mod[$\Delta\varphi(v_0)$,$2\pi$]. The frequency is tuned over 12 GHz with an interval 0.5 GHz by adjusting the cavity length of the laser. The power cost of the control light pulse is 52.4(0.6) fJ/bit. According to our experimental measurements, the extinction ratio reaches the maximum of 17.5(4.4) dB at $v_0$ and decreases when the light frequency deviates away from $v_0$. At frequency $|v - v_0| = 2.1$ GHz, the extinction ratio is half of the maximum at $v_0$. Thus, we obtain a 3 dB bandwidth of 4.2 GHz.

We also theoretically fit the experimental data with Eq. 7. In the fitting, we use $\eta_1 = 0.98$ and $\eta_2 = 0.82$. By fitting experimental data, we obtain an estimation $\Delta L = 5.5(0.1)$ mm (see the red curve in Fig. 5a). If we can reduce the OPD in experiment, the bandwidth can be considerably broadened. For example, when $\Delta L = 2$ mm, the bandwidth can be improved to 11.55 GHz (see the orange curve in Fig. 5a). Figure 5b shows the estimation of the bandwidth for different OPD $\Delta L$. The bandwidth can exceed 110 GHz when the OPD reduces to $\Delta L =$0.2 mm. In an ideal case where the effective optical path of two MZI arms matches perfectly, i.e. $\Delta L = 0$ and without considering material dispersion, the extinction ratio is independent of light frequency because our system is linear and cavity free. In this ideal case, the bandwidth can be very large. This is one of important advantages of our scheme. However, the applied chiral material also has dispersion for different light frequencies. This dispersion will further limit the available bandwidth.

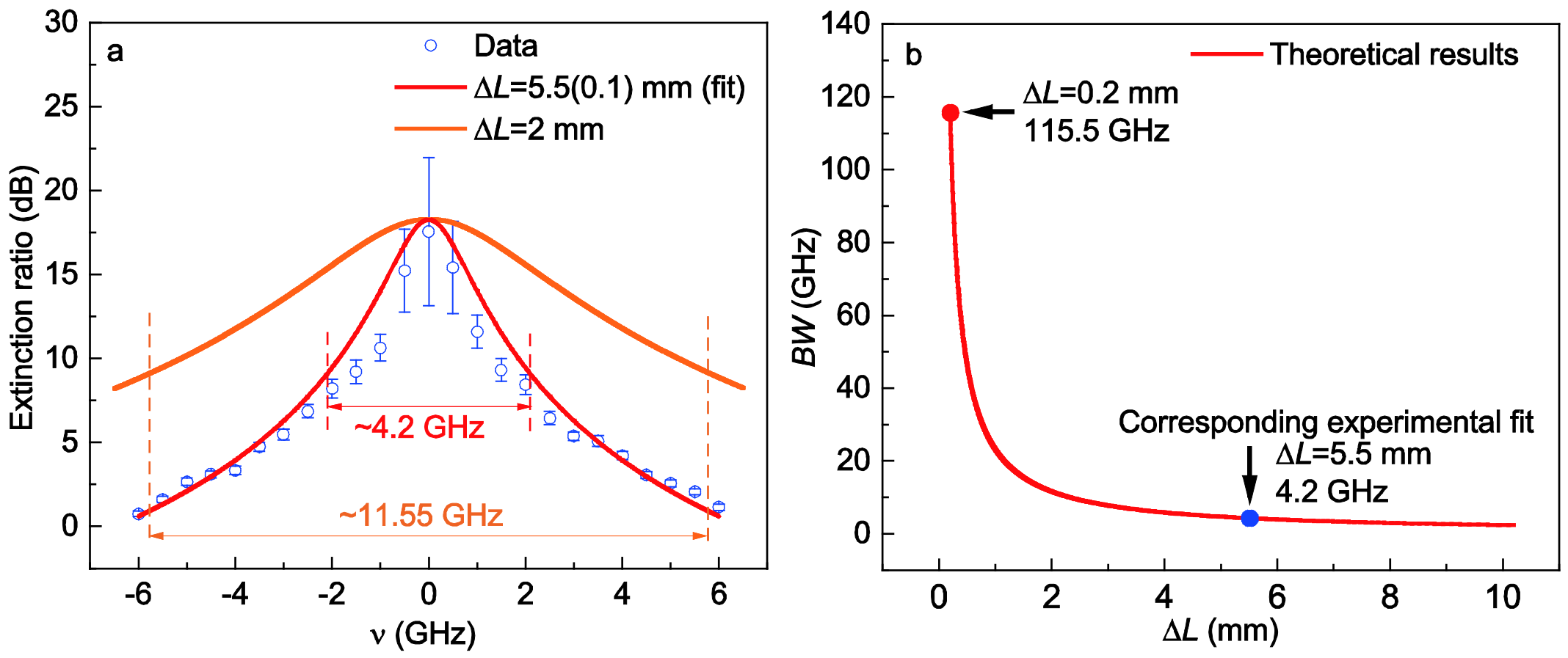

**Fig. 5 Bandwidth of all-optical switching. a** Extinction ratio as a function of laser frequency with respect to an optimal frequency 384.2782 THz where $\mathrm{mod}[\Delta\varphi, 2\pi]$ is minimal. The energy of the control light pulse is 52.4(0.6) fJ/bit, corresponding to the control power of 188.8(2.3) pW. The blue hollow circles represent experimental data of single measurement. The red curve represents the fitting result of experimental data. The orange curve represents theoretical estimation with $\Delta L = 2$ mm. **b** Theoretical estimation of bandwidth as a function of $\Delta L$.

## Discussion and Conclusion

We have reported an all-optical switching by using a chiral MZI at room temperature with an ultralow power cost of 1.3(0.1) fJ/bit and a bandwidth of 4.2 GHz. An extinction ratio of 14.7(2.8) dB is also achieved for multiple-times averaged measurement. The switching power consumption is limited by the system noise level (0.23 pW) and measuring precision of oscilloscope. The bandwidth can be improved up to 110 GHz if we can reduce the OPD in future. Our all-optical switching using a linear system has the potential to be integrated on a chip when one arm of an on-chip MZI is surrounded by a chiral material with giant optical activity[66]. Our work provides an opportunity for ultralow-power all-optical information processing and opens a door for the study of ultralow-power and ultrafast integrated photonic devices.

## Methods

**Chiral material measurement.** In our experiment, chiral material with optical activity is inserted in a MZI. Such material is circular double refractive. It introduces phase shifts $\phi_l$ and $\phi_r$ to LCP and RCP light, respectively, but $\phi_l \neq \phi_r$. The phase difference, $\Delta\phi = \phi_l - \phi_r$, is proportional to the sample length and the material rotatory power. When a linearly polarized light transmits this kind of chiral material, the polarization planar rotates with a degree of $\Delta\phi/2$.

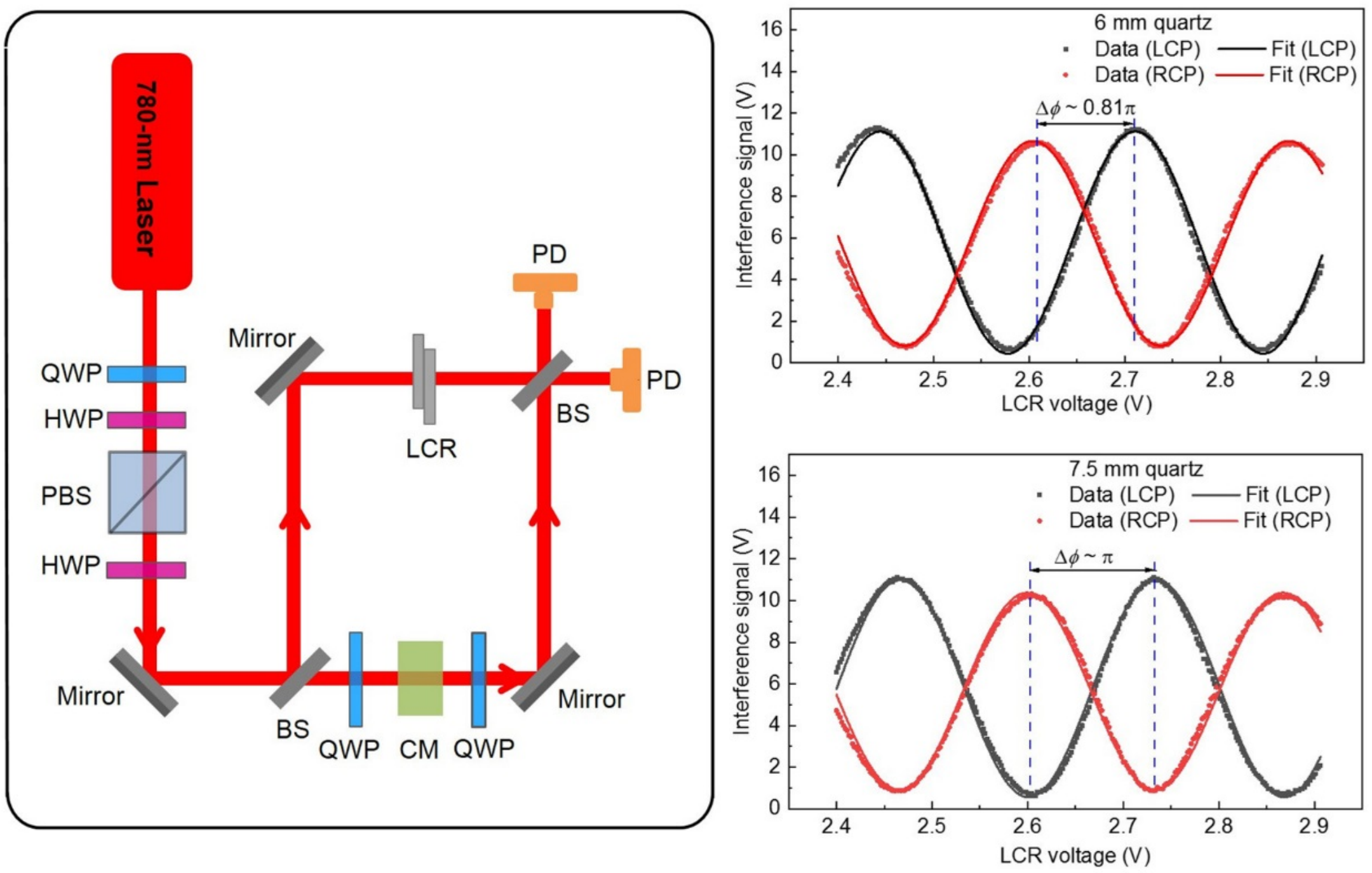

**Fig. 6 Chiral material measurement. a** Experimental set-up of optical rotation measurement. The simplified expressions of all optical devices are the same as those in Fig. 1a. **b** and **c** Experimental and fitting results for 6 mm and 7.5 mm quartz crystals, respectively. Black squares and red circles represent interference signal intensities in different LCR voltages when the lights are LCP and RCP, respectively. Black and red curves represent the best fits of experimental results.

In our all-optical switching experiment, we use a quartz crystal as chiral material. We measure the optical ration of the chiral crystal by using the setup show in Fig. 6a. After passing through QWP, HWP, PBS and HWP, the light from a tunable external cavity diode laser (Model DLC pro, Toptica Company) is prepared to V-polarized. This V-polarized laser beam then incidents on a MZI to measure the optical rotation of the chiral material. A LCR is inserted in the upper path to continuously adjust the phase difference between the two paths of MZI, while the light remains V-polarized. In contrast, the chiral material is inserted between two QWPs in the lower path. The first QWP transforms the V-polarized light into LCP or RCP. After that, the second QWP converts the LCP or RCP light back into V-polarized so that lights in two paths interfere on the second BS. When the light incident to chiral material is LCP, the interference signal is given by

$$I_{-} = \frac{I_0'}{4}\left[\eta + \eta_l - 2\sqrt{\eta\eta_l}\cos(\phi - \phi_l)\right], \tag{10}$$

where $I_0'$ is the intensity of the input light before the first BS. $\eta$ and $\eta_l$ are the transmittances of LCR and chiral material for a LCP light beam, respectively. $\phi$ and $\phi_l$ are the phases introduced by LCR and chiral material. The absorptions and phases introduced by two QWPs are constant and left out. When the light incident to chiral material is RCP, the interference signal is given by

$$I_{+} = \frac{I_0'}{4}\left[\eta + \eta_r - 2\sqrt{\eta\eta_r}\cos(\phi - \phi_r)\right], \tag{11}$$

where $\eta_r$ is the transmittance of chiral material for a RCP light beam. $\phi_r$ is the phase caused by the chiral material for a RCP light beam. Then, the optical rotation is given by

$$\alpha = \frac{\Delta\phi}{2L}, \tag{12}$$

where $\Delta\phi = \phi_l - \phi_r$. It means the optical rotatory power of chiral material per unitary length. Therefore, the optical rotation $\alpha$ can be calculated from the phase difference $\Delta\phi$ between two interference curves. We measure the interference curves of a MZI embedded with a 6 mm quartz crystal, as shown in Fig. 6b. From the phase difference $\Delta\phi = 0.81\pi$, we calculate that the optical rotation is 12°/mm. As estimation, a phase difference $\Delta\phi = \pi$ is obtained when the quartz crystal is 7.5 mm long, see Fig. 5c. This means that a 7.5 mm quartz crystal can make the polarization plane rotate $90°$. Hence, we chose a 7.5 mm quartz crystal to for our all-optical switching experiment.

**Phase modulation in chiral MZI.** To conduct a high-performance optical switching, we need to find the optimal OPD $\Delta L$ for the logic “off” state such that the effective phase difference between two arms is zero, i.e. mod($\Delta\varphi, 2\pi$)=0. We adjust this phase difference by a variable optical delay system consisting of a prism and two reflecting mirrors. The prism is fixed in a motorized positioning platform (Model KMTS50E, Thorlabs) with a minimal achievable incremental movement 50 nm.

## References

1. Espinola, R. L., Tsai, M. C., Yardley, J. T. & Osgood, R. M. Fast and low-power thermooptic switch on thin silicon-on-insulator. *IEEE Photonics Technol. Lett.* **15**, 1366–1368 (2003).
2. Lv, J. *et al.* Graphene-embedded first-order mode polymer Mach–Zender interferometer thermo-optic switch with low power consumption. *Opt. Lett.* **44**, 4606 (2019).
3. Bowley, C. C., Kossyrev, P. A., Crawford, G. P. & Faris, S. Variable-wavelength switchable Bragg gratings formed in polymer-dispersed liquid crystals. *Appl. Phys. Lett.* **79**, 9–11 (2001).
4. Komar, A. *et al.* Dynamic Beam Switching by Liquid Crystal Tunable Dielectric Metasurfaces. *ACS Photonics* **5**, 1742–1748 (2018).
5. Aplet, L. J. & Carson, J. W. A Faraday Effect Optical Isolator. *Appl. Opt.* **3**, 544 (1964).
6. Shalaby, M., Peccianti, M., Ozturk, Y. & Morandotti, R. A magnetic non-reciprocal isolator for broadband terahertz operation. *Nat. Commun.* **4**, 1–7 (2013).
7. Beck, M. *et al.* Acousto-optical multiple interference switches. *Appl. Phys. Lett.* **91**, 061118 (2007).
8. Kang, M. S., Nazarkin, A., Brenn, A. & Russell, P. S. J. Tightly trapped acoustic phonons in

photonic crystal fibres as highly nonlinear artificial Raman oscillators. *Nat. Phys.* **5**, 276–280 (2009).
9. DelRe, E., Crosignani, B., Di Porto, P., Palange, E. & Agranat, A. J. Electro-optic beam manipulation through photorefractive needles. *Opt. Lett.* **27**, 2188 (2002).
10. Haffner, C. *et al.* Low-loss plasmon-assisted electro-optic modulator. *Nature* **556**, 483–486 (2018).
11. Tsao, S.-L., Guo, H.-C. & Chen, Y.-J. Design of a 2 × 2 MMI MZI SOI electro-optic switch covering C band and L band. *Microw. Opt. Technol. Lett.* **33**, 262–265 (2002).
12. Ying, Z. & Soref, R. Electro-optical logic using dual-nanobeam Mach-Zehnder interferometer switches. *Opt. Express* **29**, 12801 (2021).
13. Reed, G. T., Mashanovich, G., Gardes, F. Y. & Thomson, D. J. Silicon optical modulators. *Nat. Photonics* **4**, 518–526 (2010).
14. Almeida, V. R., Barrios, C. A., Panepucci, R. R. & Lipson, M. All-optical control of light on a silicon chip. *Nature* **431**, 1081–1084 (2004).
15. Gopal, A. V., Yoshida, H., Neogi, A., Georgiev, N. & Mozume, T. Intersubband absorption saturation in InGaAs-AlAsSb quantum wells. *IEEE J. Quantum Electron.* **38**, 1515–1520 (2002).
16. Cong, G. W., Akimoto, R., Akita, K., Hasama, T. & Ishikawa, H. Low-saturation-energy-driven ultrafast all-optical switching operation in (CdS/ZnSe)/BeTe intersubband transition. *Opt. Express* **15**, 12123 (2007).
17. Nakamura, S., Ueno, Y. & Tajima, K. Femtosecond switching with semiconductor-optical-amplifier-based Symmetric Mach–Zehnder-type all-optical switch. *Appl. Phys. Lett.* **78**, 3929–3931 (2001).
18. Nielsen, M. L., Mørk, J., Suzuki, R., Sakaguchi, J. & Ueno, Y. Experimental and theoretical investigation of the impact of ultra-fast carrier dynamics on high-speed SOA-based all-optical switches. *Opt. Express* **14**, 331 (2006).
19. Andrekson, P. A., Sunnerud, H., Oda, S., Nishitani, T. & Yang, J. Ultrafast, atto-Joule switch using fiber-optic parametric amplifier operated in saturation. *Opt. Express* **16**, 16956 (2008).
20. Barthelemy, P. et al. Optical switching by capillary condensation. Nat. Photonics 1, 172–175 (2007).
21. Harris, S. E. Electromagnetically induced transparency. *Phys. Today* **50**, 36–42 (1997).
22. Harris, S. E. & Yamamoto, Y. Photon Switching by Quantum Interference. *Phys. Rev. Lett.* **81**, 3611–3614 (1998).
23. Yan, M., Rickey, E. G. & Zhu, Y. Observation of absorptive photon switching by quantum interference. *Phys. Rev. A* **64**, 041801 (2001).
24. Kang, H., Hernandez, G., Zhang, J. & Zhu, Y. Phase-controlled light switching at low light levels. *Phys. Rev. A* **73**, 011802 (2006).
25. Wei, X., Zhang, J. & Zhu, Y. All-optical switching in a coupled cavity-atom system. *Phys. Rev. A* **82**, 033808 (2010).
26. Braje, D. A., Balić, V., Yin, G. Y. & Harris, S. E. Low-light-level nonlinear optics with slow light. *Phys. Rev. A* **68**, 041801 (2003).
27. Chen, Y.-F., Tsai, Z.-H., Liu, Y.-C. & Yu, I. A. Low-light-level photon switching by quantum interference. *Opt. Lett.* **30**, 3207 (2005).
28. Lin, W.-H., Liao, W.-T., Wang, C.-Y., Lee, Y.-F. & Yu, I. A. Low-light-level all-optical switching based on stored light pulses. *Phys. Rev. A* **78**, 033807 (2008).

29. Jiang, W., Chen, Q., Zhang, Y. & Guo, G.-C. Optical pumping-assisted electromagnetically induced transparency. *Phys. Rev. A* **73**, 053804 (2006).
30. Bason, M. G., Mohapatra, A. K., Weatherill, K. J. & Adams, C. S. Narrow absorptive resonances in a four-level atomic system. J. Phys. B At. *Mol. Opt. Phys.* **42**, 075503 (2009).
31. Wang, H., Goorskey, D. & Xiao, M. Controlling the cavity field with enhanced Kerr nonlinearity in three-level atoms. *Phys. Rev. A* **65**, 051802 (2002).
32. Yang, X., Li, S., Cao, X. & Wang, H. Light switching at low light level based on changes in light polarization. *J. Phys. B At. Mol. Opt. Phys.* **41**, 085403 (2008).
33. Dawes, A. M. C. All-Optical Switching in Rubidium Vapor. *Science* **308**, 672–674 (2005).
34. Dayan, B. *et al.* A photon turnstile dynamically regulated by one atom. *Science* **319**, 1062–5 (2008).
35. Peyronel, T. *et al.* Quantum nonlinear optics with single photons enabled by strongly interacting atoms. *Nature* **488**, 57–60 (2012).
36. Firstenberg, O. *et al.* Attractive photons in a quantum nonlinear medium. *Nature* **502**, 71–75 (2013).
37. Chen, W. *et al.* All-Optical Switch and Transistor Gated by One Stored Photon. *Science* **341**, 768–770 (2013).
38. Stern, L., Grajower, M. & Levy, U. Fano resonances and all-optical switching in a resonantly coupled plasmonic–atomic system. *Nat. Commun.* **5**, 4865 (2014).
39. Katouf, R., Komikado, T., Itoh, M., Yatagai, T. & Umegaki, S. Ultra-fast optical switches using 1D polymeric photonic crystals. *Photonics Nanostructures - Fundam. Appl.* **3**, 116–119 (2005).
40. Aoki, T. *et al.* Efficient Routing of Single Photons by One Atom and a Microtoroidal Cavity. *Phys. Rev. Lett.* **102**, 083601 (2009).
41. Hu, X., Jiang, P., Ding, C., Yang, H. & Gong, Q. Picosecond and low-power all-optical switching based on an organic photonic-bandgap microcavity. *Nat. Photonics* **2**, 185–189 (2008).
42. McCutcheon, M. W. *et al.* All-optical conditional logic with a nonlinear photonic crystal nanocavity. *Appl. Phys. Lett.* **95**, 221102 (2009).
43. Nozaki, K. *et al.* Sub-femtojoule all-optical switching using a photonic-crystal nanocavity. *Nat. Photonics* **4**, 477–483 (2010).
44. Volz, T. *et al.* Ultrafast all-optical switching by single photons. *Nat. Photonics* **6**, 605–609 (2012).
45. Xia, K. & Twamley, J. All-Optical Switching and Router via the Direct Quantum Control of Coupling between Cavity Modes. *Phys. Rev. X* **3**, 031013 (2013).
46. Brown, A., Joshi, A. & Xiao, M. Controlled steady-state switching in optical bistability. *Appl. Phys. Lett.* **83**, 1301–1303 (2003).
47. Brown, A. W. & Xiao, M. All-optical switching and routing based on an electromagnetically induced absorption grating. *Opt. Lett.* **30**, 699 (2005).
48. Hu, X.-X. *et al.* Cavity-enhanced optical controlling based on three-wave mixing in cavity-atom ensemble system. *Opt. Express* **27**, 6660 (2019).
49. Ono, M. *et al.* Ultrafast and energy-efficient all-optical switching with graphene-loaded deep-subwavelength plasmonic waveguides. *Nat. Photonics* **14**, 37–43 (2020).
50. Xia, K. *et al.* Reversible nonmagnetic single-photon isolation using unbalanced quantum coupling. *Phys. Rev. A* **90**, 043802 (2014).
51. Tang, L. *et al.* On-chip chiral single-photon interface: Isolation and unidirectional emission.

*Phys. Rev. A* **99**, 043833 (2019).
52. Zhang, S. *et al.* Thermal-motion-induced non-reciprocal quantum optical system. *Nat. Photonics* **12**, 744–748 (2018).
53. Cao, Q.-T. *et al.* Experimental Demonstration of Spontaneous Chirality in a Nonlinear Microresonator. *Phys. Rev. Lett.* **118**, 033901 (2017).
54. Shen, Z. *et al.* Experimental realization of optomechanically induced non-reciprocity. *Nat. Photonics* **10**, 657–661 (2016).
55. Huang, R., Miranowicz, A., Liao, J.-Q., Nori, F. & Jing, H. Nonreciprocal Photon Blockade. *Phys. Rev. Lett.* **121**, 153601 (2018).
56. Lu, X., Cao, W., Yi, W., Shen, H. & Xiao, Y. Nonreciprocity and Quantum Correlations of Light Transport in Hot Atoms via Reservoir Engineering. *Phys. Rev. Lett.* **126**, 223603 (2021).
57. Xu, D. *et al.* Synchronization and temporal nonreciprocity of optical microresonators via spontaneous symmetry breaking. *Adv. Photonics* **1**, 1 (2019).
58. Hu, X.-X. *et al.* Noiseless photonic non-reciprocity via optically-induced magnetization. *Nat. Commun.* **12**, 2389 (2021).
59. Dong, M. X. *et al.* All-optical reversible single-photon isolation at room temperature. *Sci. Adv.* **7**, (2021).
60. Liang, C. *et al.* Collision-Induced Broadband Optical Nonreciprocity. *Phys. Rev. Lett.* **125**, 123901 (2020).
61. Huang, X., Lu, C., Liang, C., Tao, H. & Liu, Y.-C. Loss-induced nonreciprocity. *Light Sci. Appl.* **10**, 30 (2021).
62. Li, E.-Z. *et al.* Experimental demonstration of cavity-free optical isolators and optical circulators. *Phys. Rev. Res.* **2**, 033517 (2020).
63. Lodahl, P. *et al.* Chiral quantum optics. *Nature* **541**, 473–480 (2017).
64. Li, T., Miranowicz, A., Hu, X., Xia, K. & Nori, F. Quantum memory and gates using a Lambda-type quantum emitter coupled to a chiral waveguide. *Phys. Rev. A* **97**, 062318 (2018).
65. Li, T., Wang, Z. & Xia, K. Multipartite quantum entanglement creation for distant stationary systems. *Opt. Express* **28**, 1316 (2020).
66. Tajitsu, Y. *et al.* Huge optical rotatory power of uniaxially oriented film of poly-L-lactic acid. *J. Mater. Sci. Lett.* **18**, 1785–1787 (1999).

## Acknowledgements

This work is supported by National Key Research and Development Program of China (Grants No. 2019YFA0308700, No.2017YFA0303701, and No.2017YFA0303703), the National Natural Science Foundation of China (NSFC) (Grants No. 11874212 and No. 11890704), and the Program for Innovative Talents and Entrepreneurs in Jiangsu.

## Author contributions

Y.R., H.W., L.T., and Z.L. performed the theoretical design, numerical simulations, and experimental measurements. S.G. performed the LCR sample fabrication. Y.R., H.Z., F.X., W.H., M.X., K.X., and Y.L. contributed to the interpretation of results and participated in manuscript preparation. K.X. and Y.L. supervised the whole project.

## Additional information

Competing financial interests: The authors declare no competing financial interests.